\documentclass[11pt,a4paper]{article}
\usepackage{authblk}
\usepackage{amsmath,amssymb}
\usepackage[dvips]{graphicx}
\usepackage{array}
\usepackage{comment}
\usepackage{color}
\usepackage{psfrag}
\usepackage{soul}
\usepackage{rotating}
\usepackage{multirow}
\usepackage{enumerate}
\usepackage{caption}
\usepackage{bbm}
\usepackage{bm}
\usepackage[normalem]{ulem}
\usepackage[super]{natbib}

\newcommand{\equ}[1]{(\ref{#1})}
\newcommand{\tab}[1]{Table~\ref{#1}}

\everymath{\displaystyle}
\newcommand{\norm}[1]{\left\lVert#1\right\rVert}
\newcommand{\argmin}{\operatornamewithlimits{argmin}}

\newcommand{\clr}{\operatorname{clr}}
\newcommand{\ilr}{\operatorname{ilr}}

\newcommand{\ATS}{\operatorname{ATS}}
\newcommand{\ARL}{\operatorname{ARL}}
\newcommand{\E}{\operatorname{E}}

\title{Anomaly Detection for Compositional Data using VSI MEWMA control chart \footnotemark[1]}
\author[1,2]{Thi Thuy Van Nguyen} 
\author[2]{Cédric Heuchenne} 
\author[3]{Kim Phuc Tran} 

\affil[1]{HEC Liège - Management School of the University of Liège, Belgium}
\affil[2]{International Research Institute for Artificial Intelligence and Data Science, Dong A University, Danang, Vietnam}
\affil[3]{Univ.Lille, ENSAIT, GEMTEX, France}
\date{}

\begin{document}

\footnotetext[0]{Corresponding author: C\'{e}dric Heuchenne, Email: C.Heuchenne@uliege.be}
\footnotetext[1]{This paper was submitted to ``10th IFAC Conference on Manufacturing Modelling, Management and Control" on 14/02/2022.}
\maketitle

\begin{abstract}                % Abstract of not more than 250 words.
 In recent years, the monitoring of compositional data using control
 charts has been investigated in the Statistical Process Control field. In this study, we will design a Phase II Multivariate Exponentially Weighted  Moving Average (MEWMA) control chart with variable sampling intervals to monitor compositional data based on isometric log-ratio transformation. The Average Time to Signal will be computed based on the Markov chain approach to investigate the performance of proposed chart. We also propose an optimal procedure to obtain the optimal control limit, smoothing constant, and out-of-control Average Time to Signal for different shift sizes and short sampling intervals. The performance of proposed chart in comparison with the standard MEWMA chart for monitoring compositional data is also provided. Finally, we end the paper with a conclusion and some recommendations for future research.  
\end{abstract}

\textbf{Keyword:}  Compositional data, Markov chain, VSI-MEWMA, control chart, Data Science

%===============================================================================

\section{\textbf{INTRODUCTION}}\label{sec:intro}

In the global competitive economy nowadays, an extremely important task for manufacturing companies is to not only offer high-quality products but also reduce waste and increase efficiency in the production processes. The development of advanced technologies in Artificial Intelligence and Data Science fields makes this task more possible, but also more challenging when competing with other companies. Therefore, making a smart decision in manufacturing becomes a crucial task in any production company. In this context, early detection of abnormal products as well as assignable causes to fix the production system as soon as possible is an indispensable part, and Statistical Control Process (SPC) is one of the most effective methods to accomplish this task. Through control charts, SPC helps manufacturing companies monitor product qualities and discover the defects in the production lines. In SPC literature, many studies have been done to design a variety of control charts for monitoring different types of process data, see \cite{Montgomery2013}. Among these different data, compositional data (CoDa) are vectors whose components are strictly positive and they often present the proportions, percentages, or frequencies of some whole. Their applications can be found in many domains such as chemical research, econometrics, and the food industry, see \cite{Aitchison1986}. Due to the constraint on the sum of components of the CoDa vector, it can not be treated as normal data. 

In SPC literature, the studies in the control charts for monitoring CoDa data are still limited.  In \cite{Boyles1997}, a chi-squared type control chart for monitoring CoDa data was proposed. Recently, the authors in \cite{Vives2014a} investigated a $T^2$ control chart for monitoring CoDa with $p =3$  and then \cite{Vives2014b} extended the work in \cite{Vives2014a} for individual observations case. Two methods for interpretations of out-of-control signal of individual $T_C^2$ control chart in case $p >3$ was proposed in \cite{Vives2016}. In \cite{Tran-coda2017}, the authors proposed a MEWMA-CoDa chart for monitoring CoDa with arbitrary components. This type of control chart was shown to be effective in detecting small to moderate process shift sizes and outperforming its competitor ($T^2$-CoDa chart). The influence of measurement errors on the performance of $T^2$, MEWMA chart for monitoring CoDa were investigated in \cite{Zaidi2019} and \cite{Zaidi2020}, respectively.
In these control charts, the authors suggested using an isometric log-ration (ilr) transformation to transform CoDa to vector in $\mathbb{R}^{p-1}$ space to handle the constraint of CoDa and the average run length (ARL) to evaluate the performance of proposed control charts. 

In the control charts mentioned above, the fixed sampling interval (FSI) was supposed to use. Recently, the design of control charts tends to use variable sampling intervals (VSI). In these charts, the sampling interval between two consecutive samples is allowed to vary due to the value of the current control statistic. Many studies on the VSI control chart have been published so far, see, for example, \cite{Castagliola2013}, and \cite{Nguyen2018VSI}, among many others.  As we know, the VSI MEWMA control chart for monitoring CoDa has not been used. Consequently, in this study, we propose a VSI MEWMA type control chart for monitoring CoDa, namely VSI MEWMA-CoDa, with arbitrary components based on ilr transformation. The modification of the Markov chain approach proposed by \cite{Lee2009} will be used to compute average time to signal (ATS), criteria to access the performance of VSI control charts. 

The rest of this paper is organized as follows: In Section
\ref{sec:modeling}, the modeling of CoDa and the suggested isometric log-ratio transformation are introduced; the VSI MEWMA-CoDa control chart together with the Markov chain approach and optimization procedure to find the optimal parameters are given in Section
\ref{sec:VSIMEWMA}; in Section
\ref{sec:performance}, the performance of the VSI MEWMA-CoDa chart with different scenarios are provided; conclusions and
some recommendations for further researches are given in Section
\ref{Sec:Conclusion}.

\section{\textbf{Modeling of Compositional Data}}
\label{sec:modeling}

%In statistics, CoDa are vectors whose components are strictly %positive and they often represent proportions, percentages, %or frequencies of interested subjects. Their applications can %be found in many domains such as chemical researches, %econometric and survey data analyses, and food industry, see %\citet{Aitchison1986}. 

By definition, a row vector, $\textbf{x} = (x_1, x_2, \ldots , x_p)$, is a p-part composition when its components are strictly positive and they carry only relative information, see \cite{Aitchison1986}, \cite{Pawlowsky2015}. The relative information here refers only to the proportions between components of the composition, regardless of their numerical values. The sum of the components of $\textbf{x}$, $\sum_{i=1}^{p}x_{i}$, is a constant $\kappa$. For instance, $\kappa = 100$ refers to measurements in percentage while  $\kappa = 1$ means that the measurements are proportions. Each composition can be considered as an equivalent class made of proportional factors since the ratios between its components do not change when multiplying it by a positive constant. In this case, if $\textbf{x}, \textbf{y}$ are compositions and $\textbf{x} = \lambda \textbf{y}$  for some constants $\lambda$, we say that $\textbf{x}, \textbf{y}$ are compositionally equivalent. To check the equivalency of the two compositions, we can use the closure function $\mathcal{C}(\textbf{x})$, defined as 
 \begin{equation*}
     \mathcal{C}(\textbf{x}) = \left( \frac{\kappa \cdot x_{1}}{\sum_{i=1}^{p}x_{i}}, \ \frac{\kappa \cdot x_{2}}{\sum_{i=1}^{p}x_{i}}, \ldots,  \frac{\kappa \cdot x_{p}}{\sum_{i=1}^{p}x_{i}} \right)
 \end{equation*}
where $\kappa > 0$ is a fixed constant; in this definition, two p-part compositions $\textbf{x}, \textbf{y}$ are compositionally equivalent if $\mathcal{C}(\textbf{x})  = \mathcal{C}(\textbf{y})$.
The sample space of CoDa is the simplex,
\begin{equation*}
\begin{split}
\mathcal{S}^p = \Big\{ \textbf{x} =(x_1, x_2, \ldots , x_p) \enskip | \enskip x_{i} > 0, i = 1, \ldots, D; \\
  \sum_{i=1}^{p}x_{i} = \kappa  \Big\}
\end{split}
\end{equation*}
 In $\mathbb{R}^p$ space, we can use Euclidean geometry to add vectors or multiply vectors by scalar to obtain their properties or compute their distance. But, due to special structure of CoDa vectors in $\mathcal{S}^p$, this geometry can not be applied directly. The authors of \cite{Aitchison1986} introduced the Aitchison geometry, with two operations required for a vector space structure on $\mathcal{S}^p$:  \textit{Perturbation} and \textit{powering} operators. The perturbation $\oplus$ of $\textbf{x} \in \mathcal{S}^p$ by $\textbf{y} \in \mathcal{S}^p$ (equivalent to the addition in $\mathbb{R}^p$) is defined by 
$$\textbf{x} \oplus \textbf{y} = \mathcal{C}(x_{1}y_{1}, \ldots, x_{p}y_{p}) \in \mathcal{S}^p$$
and the powering $\odot$ of $\textbf{x} \in \mathcal{S}^p$ by a constant $\alpha \in \mathbb{R}$ (equivalent to the multiplication by a scalar operation in the $\mathbb{R}^p$) is defined by 
$$\alpha \odot \textbf{x} = \mathcal{C}(x_{1}^{\alpha}, \ldots, x_{p}^{\alpha}) \in \mathcal{S}^p$$
In practice, CoDa are often transformed to vectors in the Euclidean space to remove its constraints. The center
log-ratio (clr) transformation of vector $\textbf{x} \in \mathcal{S}^p$,  $\clr(\textbf{x})$, is an isometry from $\mathcal{S}^p$ to a subspace $U \subset \mathbb{R}^p$, defined by
\begin{align*}
        \clr(\textbf{x}) &= \left(\ln \frac{x_1}{g_{m}(\textbf{x})}, \ln \frac{x_2}{g_{m}(\textbf{x})}, \ldots, \ln \frac{x_p}{g_{m}(\textbf{x})} \right) \\
        & = (\xi_1, \xi_2, \ldots, \xi_p)
    \end{align*}
where 
\begin{equation*}
    g_{m}(\textbf{x}) = \left(\prod_{i=1}^p x_i \right)^{\frac{1}{p}} = \exp \left(\frac{1}{p}\sum_{i=1}^p x_i  \right)
\end{equation*}
 is the geometric mean of the composition and $\sum_{i=1}^p \xi_i = 0$. The inverse center
log-ratio $\clr^{-1}(\bm{\xi})$ recovering $\textbf{x}$ from $\bm{\xi} = (\xi_1, \ldots, \xi_{p})$ is 
 \begin{equation*}
     \clr^{-1}(\bm{\xi}) = \mathcal{C}(\exp(\bm{\xi})) = \mathcal{C}(\exp(\xi_1),\exp(\xi_2), \ldots, \exp(\xi_{p})).
 \end{equation*}
Egozcue et al. \cite{Egozcue2003} showed that the constraint in the component of $\clr(\textbf{x})$ makes singular the $\clr(\textbf{x})$ variance-covariance matrix for random composition. To overcome this drawback, the authors of \cite{Egozcue2003}  proposed a new transformation which is  associated with an orthogonal basis in $\mathcal{S}^p$, named isometric log-ratio (irl) transformation.  Let ${\textbf{e}_1, \textbf{e}_2, \ldots , \textbf{e}_{p-1}}$ be an orthonormal basis of $\mathcal{S}^p$. Any composition $\textbf{x} \in \mathcal{S}^p$ can be expressed as
\begin{equation*}
    \textbf{x} = \bigoplus_{i = 1}^{p-1}x_{i}^* \odot \textbf{e}_i, \enskip x_{i}^* = \langle \textbf{x},\textbf{e}_i \rangle_a = \langle \clr(\textbf{x}), \clr(\textbf{e}_i) \rangle
\end{equation*}
where $\langle , \rangle_a$ denotes the Aitchison inner product. Thus, the ilr transformation of $\textbf{x} \in \mathcal{S}^p$ is $\ilr(\textbf{x}) = \textbf{x}^* = (x_{1}^*, x_{2}^*, \ldots, x_{p-1}^*)$. 
Let $\textbf{B}$ be a $(p-1, p)$
matrix whose $i^{th}$ row is  $\clr(\textbf{e}_i), i = 1, \ldots, p-1$. This matrix is known as a contrast matrix associated with the orthonormal basis ${\textbf{e}_1, \textbf{e}_2, \ldots , \textbf{e}_{p-1}}$. The ilr transformation $\textbf{x}^*$ of composition $\textbf{x}$ can be computed by 
$$\textbf{x}^* = \ilr(\textbf{x}) =(x_1^*, \ldots, x_{p-1}^*) = \clr(\textbf{x}) \cdot \textbf{B}^{\intercal}$$
There are many candidates for an orthonormal basis in $\mathcal{S}^p$. The authors of \cite{Egozcue2005} proposed a sequential binary partition to define an orthonormal basis. In this basis, $\textbf{e}_i$ is defined to be $\mathcal{C}(e_{i,1},\ldots,e_{i,j},\ldots,e_{i,p})$ where 
\[
e_{i,j}=\left\{
  \begin{array}{ll}
    \exp{\left(\sqrt{\frac{1}{i(i+1)}}\right)} & \text{\quad if \quad} j\leq i \\
    \exp{\left(-\sqrt{\frac{i}{i+1}}\right)} & \text{\quad if \quad} j=i+1  \\
    1 & \text{\quad otherwise \quad}
  \end{array}
\right.
\]
The elements $e_{i,j}$ are called the \emph{balancing elements} of this basis. Thus, if this type of orthonormal basis is chosen to transform $\textbf{x}$, i.e., $\ilr(\textbf{x}) = \clr(\textbf{x}) \cdot \textbf{B}^{\intercal}$, the coordinates $x^*_i$ are called \emph{balances} and can be obtained by
\[
x^*_i=\sqrt{\frac{i}{i+1}}\ln\left(\frac{\left(\prod_{j=1}^ix_j\right)^{\frac{1}{i}}}{x_{i+1}}\right).
\]
From its ilr coordinate $\textbf{x}^*$, $\textbf{x}$ can be recovered by using the inverse of ilr transformation: 
\[
\ilr^{-1}(\mathbf{x}^*)=\clr^{-1}(\mathbf{x}^*\mathbf{B})
=\mathcal{C}(\exp(\mathbf{x}^*\mathbf{B})).
\]
\tab{tab:example1} illustrates the application of ilr transformation in practice for the case $p=4$. The first 4 columns present the components of 6 compositions in $\mathcal{S}^4$ and the remaining 3 columns present their corresponding ilr coordinates in $\mathbb{R}^3$. As can be seen, these $\ilr$ coordinates $x_i^*$  are not constrained any longer.
For more detail on CoDa and its properties, see \cite{Pawlowsky2015}. 
\begin{table}
\caption{An example of ilr transformation in $\mathcal{S}^4$}
  \begin{center}
    \begin{tabular}{ccccccc}
    \hline
      $x_1$ & $x_2$ & $x_3$ & $x_4$ & $x^*_1$  & $x^*_2$ & $x^*_3$ \\
      \hline
    0.10  &	0.30 &	0.50 &	0.10 &	-0.78 &	-0.87 &	0.78\\
    0.20 &	0.25 &	0.20 &	0.35 &	-0.16 &	0.09 &	-0.42\\
    0.50 &	0.10 &	0.20 &	0.20 &	1.14 &	0.09 &	0.06\\
    0.60 &	0.05 &	0.05 &	0.30 &	1.76 &	1.01 &	-0.83\\
    0.35 &	0.15 &	0.10 &	0.40 &	0.60 &	0.68 &	-0.72\\
    0.20 &	0.45 &	0.05 &	0.30 &	-0.57 &	1.46 &	-0.52\\

      \hline
    \end{tabular}  
  \end{center}
  \label{tab:example1}
\end{table}
\section{\textbf{VSI Multivariate EWMA control chart for Compositional Data}}
\label{sec:VSIMEWMA}
\subsection{\textbf{VSI MEWMA-CoDa control chart}}
Let us suppose that, at each sampling period $i=1,2,\ldots$, a sample of size $n$ independent $p$-part composition observations
$\{\mathbf{X}_{i,1},\ldots,\mathbf{X}_{i,n}\}$,
$\mathbf{X}_{i,j}\in\mathcal{S}^p$, $j=1,\ldots,n$ is collected, and suppose also that each $\mathbf{X}_{i,j}$, $j=1,\ldots,n$, follows a multivariate normal distribution $N_{\mathcal{S}^p}(\bm{\mu},\bm{\Sigma})$
on the simplex $\mathcal{S}^p$, where $\bm{\mu} \in \mathcal{S}^p$ is the center of compositions and $\bm{\Sigma}$ is their variance-covariance matrix. 
Assume that, when the process is in-control, the composition center is $\bm{\mu}_0$ and when the process is out-of-control, the composition center is $\bm{\mu}_1$.
The aim of this paper is to design a variable sampling interval MEWMA control chart (denoted by VSI-MEWMA-CoDa) to monitor the \emph{center} $\bm{\mu}$ of a $p$-part compositional process. Since CoDa data has a constant constraint on its components, the traditional VSI-MEWMA control chart may not perform well on monitoring this type of data. In \cite{Tran-coda2017}, instead of directly monitoring the composition center $\bm{\mu}$, the authors proposed to monitor the mean vector $\bm{\mu}^* = \ilr(\bm{\mu})$ using a FSI MEWMA control chart for the sample mean coordinates vector $\bar{\mathbf{X}}^*_i$. In this study, we will apply the idea in \cite{Tran-coda2017} to investigate a VSI-MEWMA control chart for monitoring a compositional process. 

Let $\{\mathbf{X}^*_{i,1},\ldots,\mathbf{X}^*_{i,n}\}$ be the 
corresponding $\ilr$ coordinates of $\{\mathbf{X}_{i,1},\ldots,\mathbf{X}_{i,n}\}$, 
i.e. $\mathbf{X}^*_{i,j}=\ilr(\mathbf{X}_{i,j})\in\mathbb{R}^{p-1}$. Since $\mathbf{X}_{i,j}$ follows a multivariate normal distribution  $N_{\mathcal{S}^p}(\bm{\mu},\bm{\Sigma})$ on $\mathcal{S}^p$, its corresponding $\ilr$ coordinate $\mathbf{X}^*_{i,j}$
follows a multivariate normal distribution
$N_{\mathbb{R}^{p-1}}(\bm{\mu}^*,\bm{\Sigma}^*)$  on $\mathbb{R}^{p-1}$, where $\bm{\mu}^*=\ilr(\bm{\mu}) \in \mathbb{R}^{p-1}$ is the
mean vector, $\bm{\Sigma}^*$ is the $(p-1,p-1)$ variance-covariance matrix of the ilr transformed data. The values of parameters
$\bm{\mu}^*$ and $\bm{\Sigma}^*$ depend on the particular choice of matrix $\mathbf{B}$ chosen in ilr transformation (see \cite{Pawlowsky2015} and section \ref{sec:modeling}). 
 Denote the ilr coordinates of  in-control composition center $\bm{\mu}_0$ and out-of-control composition center $\bm{\mu}_1$ are $\bm{\mu}_0^*$ and $\bm{\mu}_1^*$, respectively. The average of $n$ independent $p$-part compositional observations is defined by
\[
\bar{\mathbf{X}}_i=\frac{1}{n}\odot(\mathbf{X}_{i,1}\oplus\cdots
\oplus\mathbf{X}_{i,n})
\]
then its ilr coordinate $\bar{\mathbf{X}}^*_i$ is  
$\bar{\mathbf{X}}^*_i=\ilr(\bar{\mathbf{X}}_i)=\frac{1}{n}(\ilr(\mathbf{X}_{i,1})+\cdots+\ilr(\mathbf{X}_{i,n}))=\frac{1}{n}(\mathbf{X}^*_{i,1}+\cdots+
\mathbf{X}^*_{i,n}) \in \mathbb{R}^{p-1}$. \\
We first recall the FSI MEWMA-CoDa control chart proposed by the authors in \cite{Tran-coda2017} as follows. Let the MEWMA vector $\mathbf{W}_i$ be 
\[
\mathbf{W}_i=r(\bar{\mathbf{X}}^*_i-\bm{\mu}_0^*)+(1-r)\mathbf{W}_{i-1},
\,i=1,2,\ldots
\]
where $\mathbf{Y}_0=\mathbf{0}$, $r\in(0,1]$ is a fixed smoothing parameter. In FSI MEWMA-CoDa control chart, Tran et al. \cite{Tran-coda2017} suggested to monitor the statistic
\begin{equation}
\label{equ:Qi}
Q_i=\mathbf{W}_i^{\intercal}\bm{\Sigma}^{-1}_{W_i}\mathbf{W}_i,\,i=1,2,\ldots
\end{equation}
where $\bm{\Sigma}_{W_i}$ is the variance-covariance matrix of $\mathbf{W}_i$. In this work, the asymptotic form of
the variance-covariance matrix $\bm{\Sigma}_{W_i}$
\[
\bm{\Sigma}_{W_i}=\frac{r}{n(2-r)}\bm{\Sigma}^*
\]
was used to compute the plotted statistic (and it is also used in our work). An out-of-control signal is issued when
$Q_i>UCL=H$, where $H>0$ is chosen to achieve a specific value of in-control $ATS$. 

In the FSI MEWMA-CoDa control chart, the sampling interval is a fixed constant $h_F$. As for the VSI MEWMA-CoDa control chart, based on the current value of $Q_i$, the time between two successive samples $\bar{\mathbf{X}}_i, \bar{\mathbf{X}}_{i+1}$ is allowed to varied. In this chart, the control limit UCL  is held the same as in the FSI chart, and an additional warning limit $w = UWL$ $(0< UWL <UCL)$ is introduced to determine the switch between the long and short sampling intervals: The long sampling intervals $h_L$ is used when the control statistic $Q_i^2 \leq UWL^2$ (safe region) and the short sampling intervals $h_S$ is used when $UWL < Q_i^2 \leq UCL^2$ (warning region). An out-of-control signal is issued when  $Q_i^2 > UCL^2$.

\subsection{\textbf{Markov chain model}}
Suppose that the occurrence of an assignable cause makes the in-control composition center $\bm{\mu}_0$ is shifted to $\bm{\mu}_1$, or equivalently $\bm{\mu}_0^*$ is shifted to $\bm{\mu}_1^*$. In this subsection, we will discuss a method based on the Markov chain model to compute the average of the zero-state time to signal (ATS) for the VSI MEWMA-CoDa control chart. Let $ATS_0, ATS_1$ denote the ATS when the process runs in-control, and out-of-control, respectively. In comparison with other control charts, it is desirable to design a chart with smaller $ATS_1$ while their $ATS_0$ are the same. In the FSI chart, since the sampling interval $h_F$ is fixed, we have
\begin{equation*}
ATS^{\mathrm{FSI}} = h_F \times ARL^{\mathrm{FSI}}.
\end{equation*}
In the VSI chart, since the sampling interval is allowed to vary, the relation between $ATS$ and $ARL$ would be:
\begin{equation*} \label{equ:tinh Eh}
ATS^{\mathrm{VSI}} = E(h) \times ARL^{\mathrm{VSI}}.
\end{equation*}
where $E(h)$ denote the average sampling interval. 

The authors of \cite{Lowry1992} showed that the performance of a MEWMA-$\bar{X}$ chart is a function of the $n$,  $\bm{\mu}_0^*$, 
$\bm{\mu}_1^*$ and $\bm{\Sigma}^*$ only
through the non-centrality parameter $\delta$ where 
\[
\delta=\sqrt{n(\bm{\mu}_1^*-\bm{\mu}_0^*)^{\intercal}(\bm{\Sigma}^*)^{-1}
(\bm{\mu}_1^*-\bm{\mu}_0^*)}.
\]
Without loss of generality, we can assume $n=1$,
$\bm{\mu}_0^*=\mathbf{0}$ (i.e. the in-control composition center is
$\bm{\mu}_0=(\frac{1}{p},\frac{1}{p},\ldots,\frac{1}{p})$) and
$\bm{\Sigma}^*=\mathbf{I}_{p-1}$ (the identity matrix in
$\mathbb{R}^{p-1}$). In this case, the statistic $Q_i$ in
\equ{equ:Qi} is modified to $Q_i=b\norm{W_i}_2^2$ with $b=\frac{2-r}{r}$. Consequently, the control limits UCL and UWL of VSI MEWMA-CoDa are modified  to be
\[
UCL = \sqrt{H/b}, \quad UWL = \sqrt{w/b}
\]
To calculate the in- and out-of-control ATS of the
VSI MEWMA-$\bar{X}$ chart, the author of \cite{Lee2009} modified the Markov chain approach proposed in \cite{Runger1996} to approximate its calculation based on the statistic $q_i=\norm{W_i}_2$. 

Concerning the \emph{in-control} case, the one dimensional Markov chain can be used to approximate ATS.
In this case, the interval $[0,UCL']$, where $UCL'=\sqrt{H/b}$, is divided into $m+1$
sub-intervals (states): the first sub-interval has length
$\frac{g}{2}$ and the others have length $g$, where
$g=\frac{2UCL'}{2m+1}$. The probability of transition from state $i$ to state $j$, denoted by $p(i,j)$, is given by
\begin{itemize}
\item for $i=0,1,\ldots,m$ and $j=1,2,\ldots,m$,
\begin{equation*}
\begin{split}
p(i,j)=P\bigg(\Big(\frac{(j-0.5)g}{r}\Big)^2<
    \chi^2(p-1,c) \\
 < \Big(\frac{(j+0.5)g}{r}\Big)^2\bigg)
\end{split}
\end{equation*}
where $\chi^2(p-1,c)$ denotes a non central chi-square random variable with $p-1$ degrees of freedom and non-centrality parameter $c=\left(\frac{(1-r)ig}{r}\right)^2$,
\item for $j=0$,
  \[
  p(i,0)=P\left(\chi^2(p-1,c)<\left(\frac{g}{2r}\right)^2\right).
  \]
\end{itemize}
Let $\textbf{P}_1$ denote the $(m+1,m+1)$ transition probability matrix corresponding to the transient states with the elements $p(i,j)$ then the zero-state in-control $ATS$ of the VSI MEWMA-CoDa control chart is obtained by
\[
ATS=\mathbf{s}^{\intercal}(\mathbf{I}_{m+1}-\mathbf{P}_1)^{-1}\mathbf{h},
\]
where $\mathbf{s}$ is the $(m+1)$-starting probability vector,
i.e. $\mathbf{s}=(1,0,0,\ldots,0)^{\intercal}$, $\mathbf{h}$ is the $(m+1)$-vector of sampling interval with the $i^{th}$ component $h_i$ is defined by 
\begin{align*}
    h_i= \begin{cases}
   h_L ~~\text{if}~~ ig \leq UWL\\
   h_S ~~\text{if}~~ ig > UWL
    \end{cases}.
\end{align*}
The expected sampling interval $E(h)$ is calculated by 
\begin{equation*} %\label{equ:eh2}
E(h)=\frac{\mathbf{s}^{\intercal}(\mathbf{I}_{m+1}-\mathbf{P}_1)^{-1}\mathbf{h}}{\mathbf{s}^{\intercal}(\mathbf{I}_{m+1}-\mathbf{P}_1)^{-1}\mathbf{1}_{m+1}},
\end{equation*}
where $\mathbf{1}_{m+1}=(1,1,\ldots,1)^{\intercal}$ is the $m+1$ column vector of 1's. 

To calculate the zero-state ATS of VSI MEWMA-CoDa chart in the \emph{out-of-control} case, The author of \cite{Lee2009} modified the two
dimensional Markov chain approach which is originally proposed in \cite{Runger1996}. In this approach,  $\mathbf{W}_i\in\mathbb{R}^{p-1}$ is partitioned into $W_{i1}\in\mathbb{R}$ with mean $\delta \neq 0$ and
$\mathbf{W}_{i2}\in\mathbb{R}^{p-2}$ with zero mean. Then, 
$q_i=\norm{\mathbf{W}_i}_2=\sqrt{W^2_{i1}
+\mathbf{W}^{\intercal}_{i2}\mathbf{W}_{i2}}$. 

The transition probability $h(i,j)$ of $W_{i1}$ from state $i$ to state $j$ is used to analyze the out-of-control component. Applying the Markov chain-based approach with the number of states of the Markov chain is $2m_1+1$, for $i, j = 1, 2,\ldots, 2m_1+1$, we have

\begin{align*}
    h(i,j)= &\Phi\left(\frac{-UCL'+jg_1-(1-r)c_i}{r}-\delta\right) \\
& -\Phi\left(\frac{-UCL'+(j-1)g_1-(1-r)c_i}{r}-\delta\right)
\end{align*}

where $\Phi$ denotes the cumulative standard normal distribution function, $c_i=-UCL'+(i-0.5)g_1$ is the center point of state $i$ with the width of each state $g_1=\frac{2UCL'}{2m_1+1}$. 

Concerning $\mathbf{W}_{i2}$ component, the transition probability $v(i,j)$ from state $i$ to state $j$ is used to analyze the in-control component. In this case, the Markov chain approach as in in-control case will be applied
with $p-2$ replacing $p-1$. The control region is partitioned into $m_2+1$ sub-intervals (states) with the width of each states is $g_2=\frac{2UCL'}{2m_2+1}$. The transition probability $v(i,j)$ is given as follows
\begin{itemize}
\item for $i=0,1,2,\ldots,m_2$ and $j=1,2,\ldots,m_2$
\begin{equation*}
\begin{split}
v(i,j)=P\bigg(\Big(\frac{(j-0.5)g_2}{r}\Big)^2<
    \chi^2(p-2,c) \\
 < \Big(\frac{(j+0.5)g_2}{r}\Big)^2\bigg),
\end{split}
\end{equation*}
\noindent  
where $c=\left(\frac{(1-r)ig_2}{r}\right)^2$,
\item for $j=0$,
\[
v(i,0)=P\left(\chi^2(p-2,c)<\left(\frac{g_2}{2r}\right)^2\right)
\]
\end{itemize}
Let $\mathbf{H}$ denote the $(2m_1+1,2m_1+1)$ transition probability matrix of $W_{i1}$ with elements $h(i,j)$, $\mathbf{V}$ denote the
$(m_2+1,m_2+1)$ transition probability matrix of
$\norm{\mathbf{Y}_{i2}}_2$ with elements $v(i,j)$, and
$\mathbf{P}_2$ denote the transition probability matrix of two dimensional Markov chain. Since $W_{i1}$ and
$\mathbf{Y}_{i2}$ are independent, we have $\mathbf{P}_2=\mathbf{H}\otimes\mathbf{V}$,
where $\otimes$ is the Kronecker's matrices product. Matrix
$\mathbf{P}_2$ will consist of the transition probabilities of all transient and some absorbing states of the Markov chain. 

Let $\mathbf{T}$ be the $(2m_1+1,m_2+1)$- matrix with element $T(\alpha, \beta)$ given by
\[
\mathbf{T}(\alpha, \beta)=\left\{
  \begin{array}{ll}
    1& \mbox{if state ($\alpha, \beta$) is transient} \\
    0& \mbox{otherwise}
  \end{array}
\right.
\]
and $\mathbf{P}$ be the transition probability matrix containing only transient states of the Markov chain. Then, we have
$\mathbf{P}=\mathbf{T}(\alpha,\beta)\circledast\mathbf{P}_2$ where symbol $\circledast$ indicates the element-wise multiplication of matrices. 

Let $\mathbf{h}$ be the $(2m_1+1) \cdot (m_2+1)$ vector of sampling intervals for the bivariate chain. M.H. Lee \cite{Lee2009} defined $\mathbf{h}$ to be
\begin{equation*}
    \begin{split}
        \mathbf{h}^{\intercal} = \big((1,0), \ldots, (1,m_2),  (2,0), \ldots, (2,m_2), \ldots, \\
                 \ldots, (2m_1+1,0), \ldots, (2m_1+1,m_2)\big)
    \end{split}
\end{equation*}
with the element $\mathbf{h}(i,j)$ defined by
\[
\mathbf{h}(i,j)=\left\{
  \begin{array}{ll}
    h_L& \mbox{if} \enskip a_{i,j} \leq UWL^2 \\
    h_S& \mbox{if} \enskip UWL^2 < a_{i,j} \leq UCL^2 \\
    0 & \mbox{otherwise}
  \end{array}
\right.
\]
where $a_{i,j} = (i-(m_1+1))^2g_1^2 +j^2g_2^2.$

Thus, the zero-sate out-of-control $ATS$ of VSI MEWMA-CoDa control chart is
defined by $ATS=\mathbf{s}^{\intercal}(\mathbf{I}-\mathbf{P})^{-1}
\mathbf{h}$ where $\mathbf{s}$ is the initial probability vector with the component corresponding to
state $(\alpha,\beta)=(m_1+1,0)$ is equal to one and all other components are equal to zero. In case $m_1=m_2=m$, M.H. Lee \cite{Lee2006} showed that the entry corresponding to the component with value equal to 1 of $\mathbf{s}$ is the $(m(m+1)+1)$th entry. Concerning the performance of
the program used for the computation of the $ATS$, we follow the recommendation in \cite{Tran-coda2017} and decide to use $m_1=m_2=30$.
\subsection{\textbf{Optimization procedure}}
Assume that the fixed sampling interval in FSI control charts is to be a time unit, i.e. $h_F=1$. Hence, $ATS_0^{FSI} = ARL_0$. In order to evaluate the performances of VSI MEWMA-CoDa with its FSI version, we can compare their out-of-control $ATS_1$ while constraining the same in-control values of both $ATS_0$ and $E_0(h)$ (average sampling interval). Thus, the VSI MEWMA-CoDa control chart can be designed by finding the optimal combination of parameters that minimize the out-of-control $ATS_1$ subject to the predefined constraint of $ATS_0$ and $E_0(h)$. 

In general, a fixed couple $(h_S, h_L)$ is typically used, which can be chosen from the suggested list as in \cite{Castagliola2013}. However, as discussed in the study \cite{Nguyen2018VSI}, while $h_S$ is quite reasonable to fix, it seems not practical to fix $h_L$ due to the fact that when the control statistic falls into the central region, the process is still in safe and the next sampling interval can be flexible to choose if it does not influence the performance of the chart. Based on this reason, we follow the suggestion in \cite{Nguyen2018VSI} to fix the proportion between the UCL and UWL values. Let $R$ be the number such that $ UWL = R \cdot UCL$. When the control limit UCL is determined, the warning limit UWL can be computed based on the value of $R$.

Thus, the optimal design of the VSI MEWMA-CoDa control chart will consist of searching the optimal parameters $(r, H, h_L)$ which minimize the out-of-control $ATS_1$ for given shift $\delta$ subject to constraints in the in-control $ATS_0$ and $E_0(h) = 1$, i.e,

\begin{equation*} 
 (r^{*}, H^{*},h_L^{*})=\underset{(r,H,h_L)}\argmin \enskip \ATS(n, r, H, R, p-1, \delta, h_L, h_S)
\end{equation*}
subject to the constraint

\begin{align*} \label{equ:optim up 2}
    \begin{cases}
      \ATS(n, r^*, H^*, R, p-1, \delta = 0, h_L^*, h_S) = \ATS_0\\
      \E_0(h)=1
    \end{cases}
\end{align*}

By fixing the in-control predefined $ATS_0$ value, these optimal parameters can be obtained by using the two-steps optimization procedure as follows
\begin{enumerate}
\item Find the set of triples $(r,H, h_L)$ such that the in-control $ATS = ATS_0$
  and $E_0(h)=1$.
\item Among these feasible triples $(r,H, h_L)$, choose $(r^*,H^*)$ which provides the smallest out-of-control $ATS$ value for a particular shift $\delta$ in vector $\bm{\mu}_0^*$.
\end{enumerate}
As noted in \cite{Tran-coda2017}, the value of
$r$ must not be too small to avoid unreliable results and the diverging ability in the Markov Chain approach. In this paper, we fix the minimal bound to search for the smoothing parameter $r$ to be $0.05$, as recommended in many studies, including \cite{Tran-coda2017}.

\section{\textbf{PERFORMANCE OF THE VSI MEWMA-CoDa CONTROL CHART}} \label{sec:performance}

In this section, we will compare the performance of the VSI MEWMA-CoDa chart with the FSI MEWMA-CoDa chart proposed by \cite{Tran-coda2017}. The comparison will be based on the values of out-of-control $\ATS_1$ while constraining on the same in-control values of both $\ATS_0$ and $\E_0(h)$. To take advantage of the results from the study of \cite{Tran-coda2017},  save the calculation costs, and simplify the application in practice, we propose to find the near-optimal values to the VSI MEWMA-CoDa control chart as follows: 
\begin{itemize}
    \item For each optimal couple $(r^*, H^*)$ in Table 2 in study of \cite{Tran-coda2017}, the value of UWL and $h_L$ are chosen to achieve predefined $\ATS_0$ and $\E_0(h)$,
    \item After obtaining UWL and $h_L$, together with the corresponding $(r^*, H^*)$, we compute the $\ATS_1$  of VSI MEWMA-CoDa for specific shift sizes $\delta$ and compare them with $\ARL_1$ of FSI MEWMA-CoDa chart (Table 3 in \cite{Tran-coda2017}).
\end{itemize}
The procedure to find the near-optimal values is implemented based on following scenarios:
\begin{itemize}
\item $n = 1, p = 3$, $\ATS_0 = 200$, and $\E_0(h) =1$;
\item $\delta \in \{0.25, 0.50, 0.75, 1.00, 1.25, 1.50, 1.75, 2.00\}$;
\item $h_S\in \{0.1, 0.5\}$.
\end{itemize}
 The  values $\ARL_1$ (FSI column) of MEWMA-CoDa and $\ATS_1$ (VSI columns) for some different scenarios are shown in Table \ref{tab:compare}. The values of $w$ such that $UWL = \sqrt{w/b}$ and $h_L$ to obtain the near-optimal value are also provided for each scenario. Some remarks can be drawn from this results as follows
\begin{itemize}
\item The VSI MEWMA-CoDa control chart always outperforms the FSI MEWMA-CoDa control chart in detecting the process shifts. For example, when $\delta = 0.25, h_S = 0.1$, we have $\ARL_1 = 64.6$ for FSI MEWMA-CoDa chart and $\ATS_1 = 56.8$ for VSI MEWMA-CoDa chart, 
\item The VSI MEWMA-CoDa charts with smaller $h_S$ $(h_S = 0.1)$ perform better than the ones with larger  $h_S$ $(h_S = 0.5)$. For example,  when $\delta = 0.5$, we have $\ATS_1 = 19.9$ in case  $h_S = 0.1$ and $\ATS_1 = 23.5$ in case  $h_S = 0.5$,
\item When the shift sizes $\delta$ are large $(\delta \geq 1.75)$, the performance of VSI MEWMA-CoDa chart are still better than FSI MEWMA-CoDa chart, but not much.
\end{itemize}
 %In fact, it can be seen that the performance of the
 %VSI MEWMA-CoDa control chart for monitoring a $p$-part normal composition is the same as of VSI MEWMA chart for monitoring multivariate normal data with $(p-1)$ variables. 
\begin{table} 
\caption{Comparison between VSI MEWMA-CoDa and FSI MEWMA-CoDa charts}
\centering
\begin{tabular}{|c|c|c|c|c|c|}
%\toprule
\multirow{2}{*}{$\delta$} & \multirow{2}{*}{FSI} & \multicolumn{2}{c|}{$h_S=0.1$} & \multicolumn{2}{c|}{$h_S=0.5$}  \\
 \cline{3-6}
   &   & $(w, h_L)$ &  VSI& $(w, h_L)$ & VSI    \\  
   \hline
%   \midrule

0.25 & 64.6 & $(1.7, 1.6)$ &  56.8 & $(  0.7,   2.1)$ & 63.5 \\
0.50 & 26.4  & $(1.7, 1.6)$ &19.9 & $(0.9,   1.8)$ &  23.5 \\
0.75 & 15.1  &$(1.6, 1.7)$ &10.4 & $(1.0,   1.8)$ & 12.9\\
1.00 & 9.9 &$(2.9, 1.3)$ &   6.9 & $(0.9,   1.8)$ &  8.4 \\
1.25 & 7.1 & $(1.6, 1.8)$ & 4.9 & $(0.9,   2.0)$ &  6.3\\
1.50 &5.4 & $(3.5, 1.2)$ &  3.7 & $(0.9,   1.9)$ &  4.8\\
1.75 & 4.3 & $(3.7, 1.2)$ &  3.0 & $(0.8,   2.1)$ &   4.2\\
2.00 & 3.5 & $(3.6, 1.2)$ & 2.4 & $(1.1,   1.8)$ &  3.3\\                     

%\bottomrule
\end{tabular}
\label{tab:compare}
\end{table}

\section{\textbf{Conclusion}}
\label{Sec:Conclusion}
In this paper, we proposed a VSI MEWMA-CoDa control chart to
monitor a normal multivariate random vector defined as the inverse isometric log-ratio of a $p$-part composition. The optimal procedure to compute the optimal triple $(r^*,H^*, h_L^*)$ and the $\ATS$ values of the proposed
chart for different shift sizes were presented. We also proposed a method to find the near-optimal values for the VSI MEWMA-CoDa chart to utilize the results in the study of \cite{Tran-coda2017} and reduce the computation costs. The numerical performance comparison between the VSI MEWMA-CoDa chart and standard (FSI) MEWMA-CoDa control chart in terms of $\ATS_1$ (based on the near-optimal values method) showed that the VSI MEWMA-CoDa chart always outperforms the standard chart. Future research on monitoring CoDa could be concentrated on
the extension of the VSI MEWMA-CoDa chart to the VSI MCUSUM-CoDa chart, or investigating the effect of measurement
error on these charts. The methods to transform CoDa into normal data before designing these controls charts are also worthy to focus. Due to the wide applications of CoDa in the real-life, the online monitoring of CoDa should be worthy of consideration by researchers in the SPC field.

%\bibliography{SPC_Reference}
%\bibliographystyle{unsrtnat}
\end{document}